\documentclass[twoside,prb,letterpaper,preprintnumbers,superscriptaddress,notitlepage]{revtex4}

\usepackage{mathptmx}
\usepackage{helvet}
\usepackage{graphicx,fancyhdr}
\usepackage{dashbox,color}%
\usepackage{bm}
\usepackage{sidecap}%
\definecolor{lightblue}{rgb}{0.6,0.9,1}
\definecolor{myrefblue}{rgb}{0.1,0.6,1}
\definecolor{myblue}{rgb}{0,0,0}
\definecolor{nmat}{rgb}{0.7,0.04,0.26}

\newcommand{\sfcaopt}{\mbox{SmFe$_{\mathsf{0.92}}$Co$_{\mathsf{0.08}}$AsO}}

\begin{document}
\pagestyle{fancy}

\renewcommand{\headrule}{\vskip-3pt\hrule width\headwidth height\headrulewidth \vskip-\headrulewidth}

\fancypagestyle{plainfancy}{%
\lhead{}
\rhead{}
\chead{}
\lfoot{}
\cfoot{}
\rfoot{\bf\scriptsize\textsf{\thepage}}
\renewcommand{\headrulewidth}{0pt}
\renewcommand{\footrulewidth}{0pt}
}

\fancyhead[LE,RO]{}
\fancyhead[LO,RE]{}
\fancyhead[C]{}
\fancyfoot[LO,RE]{}
\fancyfoot[C]{}
\fancyfoot[LE,RO]{\bf\scriptsize\textsf{\thepage}}

\renewcommand\bibsection{\section*{\sffamily\bfseries\normalsize {References}\vspace{-10pt}\hfill~}}
\newcommand{\mysection}[1]{\section*{\sffamily\bfseries\normalsize {#1}\vspace{-10pt}\hfill~}}
\renewcommand{\subsection}[1]{\noindent{\bfseries\normalsize #1}}
\renewcommand{\bibfont}{\fontfamily{ptm}\footnotesize\selectfont}
\renewcommand{\figurename}{Figure}
\renewcommand{\refname}{References}
\renewcommand{\bibnumfmt}[1]{#1.}

\makeatletter
\long\def\@makecaption#1#2{%
  \par
  \vskip\abovecaptionskip
  \begingroup
   \small\rmfamily
   \sbox\@tempboxa{%
    \let\\\heading@cr
    \textbf{#1\hskip1pt$|$\hskip1pt} #2%
   }%
   \@ifdim{\wd\@tempboxa >\hsize}{%
    \begingroup
     \samepage
     \flushing
     \let\footnote\@footnotemark@gobble
     \textbf{#1\hskip1pt$|$\hskip1pt} #2\par
    \endgroup
   }{%
     \global \@minipagefalse
     \hb@xt@\hsize{\hfil\unhbox\@tempboxa\hfil}%
   }%
  \endgroup
  \vskip\belowcaptionskip
}%
\makeatother

\thispagestyle{plainfancy}

\fontfamily{helvet}\fontseries{bf}\selectfont
\mathversion{bold}
\begin{widetext}
\begin{figure}
\vskip0pt\noindent\hskip-0pt
\hrule width\headwidth height\headrulewidth \vskip-\headrulewidth
\hbox{}\vspace{4pt}
\hbox{}\noindent\vskip10pt\hbox{\noindent\huge\sffamily\textbf{Interaction-induced singular Fermi surface in a}}\vskip0.05in\hbox{\noindent\huge\sffamily\textbf{high-temperature oxypnictide superconductor}}
\vskip10pt
\hbox{}\noindent\begin{minipage}{\textwidth}\flushleft
\renewcommand{\baselinestretch}{1.2}
\hskip-10pt\large\sffamily A.~Charnukha$^{1,2,\dagger}$, S.~Thirupathaiah$^{1,3}$, V.~B.~Zabolotnyy$^{1,4}$, B.~B\"uchner$^{1}$, N.~D.~Zhigadlo$^{5}$, B.~Batlogg$^{5}$, A.~N.~Yaresko$^{6}$ \&~S.~V.~Borisenko$^{1}$
\end{minipage}
\end{figure}
\end{widetext}

\begin{figure}[!h]
\begin{flushleft}
{\footnotesize\sffamily
$^1$Leibniz Institute for Solid State and Materials Research, IFW, 01069 Dresden, Germany, $^2$Physics Department, University of California--San~Diego, La Jolla, CA 92093, USA, $^3$Solid State and Structural Chemistry Unit, Indian Institute of Science, Bangalore--560 012, India, $^4$Physikalisches Institut und R\"ontgen Center for Complex Materials Systems, Universit\"at W\"urzburg, 97074 W\"urzburg, Germany, $^5$Laboratory for Solid State Physics, ETH Zurich, CH-8093 Zurich, Switzerland, $^6$Max Planck Institute for Solid State Research, 70569 Stuttgart, Germany. $^\dagger$ E-mail: acharnukha@ucsd.edu.}
\end{flushleft}
\end{figure}

\fontsize{8pt}{8pt}\selectfont
\renewcommand{\baselinestretch}{0.9}
\noindent\sffamily\bfseries{In the family of iron-based superconductors, LaFeAsO-type materials possess the simplest electronic structure due to their pronounced two-dimensionality. And yet they host superconductivity with the highest transition temperature $T_{\mathrm{c}}\approx 55\ \textrm{K}$. Early theoretical predictions of their electronic structure revealed multiple large circular portions of the Fermi surface with a very good geometrical overlap (nesting), believed to enhance the pairing interaction and thus superconductivity. The prevalence of such large circular features in the Fermi surface has since been associated with many other iron-based compounds and has grown to be generally accepted in the field. In this work we show that a prototypical compound of the 1111-type, \sfcaopt, is at odds with this description and possesses a distinctly different Fermi surface, which consists of two singular constructs formed by the edges of several bands, pulled to the Fermi level from the depths of the theoretically predicted band structure by strong electronic interactions. Such singularities dramatically affect the low-energy electronic properties of the material, including superconductivity. We further argue that occurrence of these singularities correlates with the maximum superconducting transition temperature attainable in each material class over the entire family of iron-based superconductors.}
\mathversion{normal}
\normalfont\normalsize

A seemingly good qualitative agreement between the early experimental determination of the low-energy electronic band structure of the iron-based compounds and the predictions of {\it ab initio} theoretical calculations~\cite{Mazin_NatureInsights_2010,Shen_LaFePO_ElStr_2008} has shaped our understanding of the essential microscopics of these superconductors. According to this view, their Fermi surface consists of multiple large nearly circular sheets formed by holelike and electronlike bands at the center and in the corner of the Brillouin zone, respectively. Observation of a resonance peak in the inelastic neutron scattering~\cite{Osborn_INS_BKFA_2008} has supported theoretical proposals that a very good geometrical overlap between these Fermi-surface sheets of different electronic character strongly enhances electronic interactions at the wave vector connecting the center and the corner of the Brillouin zone and gives rise to a superconducting energy gap with different sign on these sheets (the so-called extended $s$-wave symmetry)~\cite{PhysRevLett.101.057003}. We employed high-resolution angle-resolved photoemission spectroscopy to demonstrate that, even though the extended $s$-wave character of the superconducting order parameter in the iron-based compounds is supported by a large body of experimental data~\cite{Johnston_Review_2010}, the original premise of two or more large nearly circular sheets of the Fermi surface is false. We show here that the Fermi surface of a prototypical 1111-type \sfcaopt\ deviates markedly from this general expectation and is highly singular, formed by multiple bands terminating at or in the immediate vicinity of the Fermi level both at the center and in the corner of the Brillouin zone and nevertheless connected by the $(\pi,\pi)$ nesting vector. These electronic singularities are, in fact, already present in the calculated electronic structure of iron-based superconductors far away from the Fermi level and are pulled to it by strong electronic interactions in the presence of a pronounced particle-hole asymmetry~\cite{Cappelluti2010S508}. It has been predicted that the presence of such singularities may lead to a significant enhancement of superconductivity~\cite{0953-2048-22-1-014004}.

\begin{figure*}[!t]
\includegraphics[width=\textwidth]{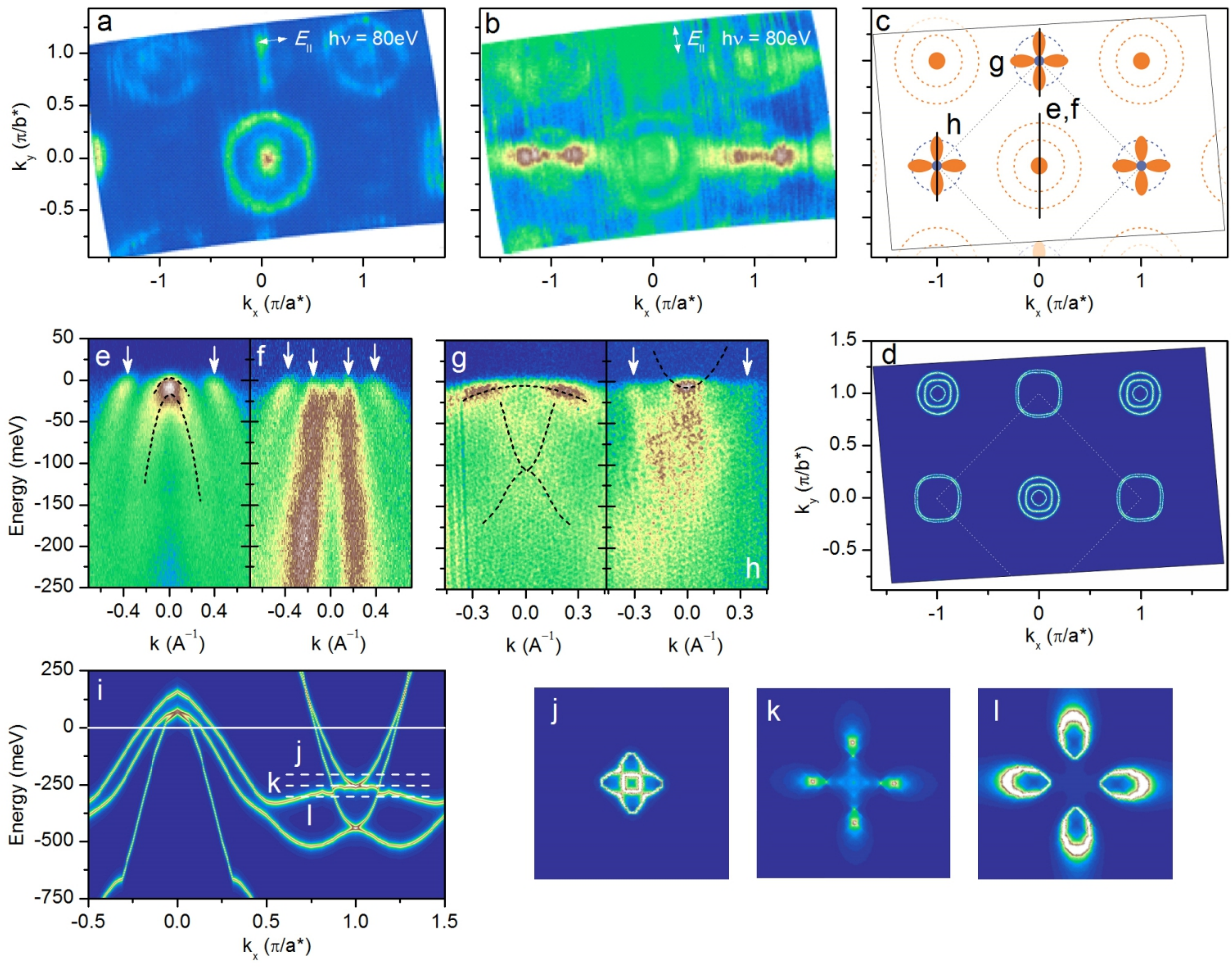}
\caption{\label{fig:fs}\textbf{Photoemission intensity at the Fermi level and low-energy electronic structure of \sfcaopt.}~\textbf{a,b,} Constant-energy maps obtained by integrating the photoemission intensity in a small energy window of $10\ \textrm{meV}$ around the Fermi level recorded at $T=1\ \textrm{K}$ using photons with an excitation energy of $80\ \textrm{eV}$ linearly polarized within (horizontal polarization) and perpendicular to (vertical polarization) the plane of incidence, respectively. The lattice parameters of the rotated by $45^\circ$ unit cell are defined as $a^*=b^*=a/\sqrt{2}$, where $a$ is the lattice parameter of the real tetragonal unit cell~\cite{PhysRevB.86.214509}.~\textbf{c,}~Schematic illustration of the experimentally observed features. Orange (blue) contours and areas indicate photoemission intensity coming from dispersions of holelike (electronlike) character. Dashed lines depict contribution from bands clearly crossing the Fermi level, while shaded areas indicate photoemission intensity from bands with their band edge located in the vicinity of the Fermi level. Gray dashed square delimits the first two-iron Brillouin zone.~\textbf{d,} Simulated photoemission intensity at the Fermi level based on the electronic structure obtained in our {\it ab initio} calculations.~\textbf{e--h,} Energy-momentum cuts along the lines shown in panel~\textbf{c} obtained using horizontal (\textbf{e,g}) and vertical (\textbf{f,h}) linear polarization of incident light. Band dispersions giving rise to the large circular features at the center and in the corner of the Brillouin zone in \textbf{a}--\textbf{c} are shown with white arrows. The other clearly observed elements of the low-energy electronic structure are indicated with dashed lines.~\textbf{i,} Spectral function along the high-symmetry $\Gamma$--M direction based on the same {\it ab initio} calculations as in panel~\textbf{d}.~\textbf{j}--\textbf{l,} Constant-energy contours around the M point at the energies indicated by the corresponding white horizontal lines in panel~\textbf{i}.}
\end{figure*}

Observation of this singular Fermi surface topology was made possible by the combination of high-quality \sfcaopt\ single crystals, which can only be grown by the laborious high-pressure technique with linear dimensions on the order of $300\ \mu\textrm{m}$ (Ref.~\onlinecite{PhysRevB.86.214509}), and a synchrotron-based angle-resolved photoemission spectroscope with a smaller beam spot, operating below $900\ \textrm{mK}$ with $<4\ \textrm{meV}$ total energy resolution~\cite{Borisenko1cubedARPES_SRN2012,Borisenko1cubedARPESJove} (see Methods). To shed light on the effect of interactions on the Fermi-surface topology of this material we have further carried out {\it ab initio} density-functional calculations of its electronic structure in the local-density approximation (see Methods).

A typical distribution of the photoemission intensity at the Fermi level in the optimally doped \sfcaopt\ is shown in Figs.~\ref{fig:fs}a,b and summarized schematically in Fig.~\ref{fig:fs}c. Due to the photoemission selection rules the intensity maps were recorded using synchrotron radiation linearly polarized both within (horizontal polarization, Fig.~\ref{fig:fs}a) and perpendicular (vertical polarization, Fig.~\ref{fig:fs}b) to the plane of incidence in order to reveal all important details of the electronic structure. Two sets of qualitatively different features are immediately visible: large circular intensity distributions both at the center and in the corner of the Brillouin zone (indicated with dashed lines in Fig.~\ref{fig:fs}c) as well as a localized intensity spot at the center and a complex propellerlike distribution of small features in the corner of the Brillouin zone (indicated with shaded areas in Fig.~\ref{fig:fs}c; due to the polarization dependence of the photoemission matrix element only two blades of the propellerlike structure are visible in photoemission maps in Figs.~\ref{fig:fs}a,b obtained with each linear polarization). The former set of features does indeed look similar to the Fermi surface predicted by our {\it ab initio} calculations of the electronic structure shown in Fig.~\ref{fig:fs}d and reported previously~\cite{2009PhyC469614M}. The localized intensity spots have a more complex character and require detailed analysis.

\begin{figure}[!t]
\includegraphics[width=\textwidth/2]{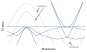}
\caption{\label{fig:bulksummary}\textbf{Extracted bulk low-energy electronic band structure of \sfcaopt.}~Schematic illustration of the bulk low-energy electronic structure of~\sfcaopt\ (blue solid lines) as well as the contribution to the photoemission signal from the polar surface (grey dashed lines). The horizontal solid black line indicates the location of the Fermi level.}
\end{figure}

To that end, we compare the low-energy electronic structure at the center ($\Gamma$ point) and in the corner (M point) of the Brillouin zone along the cuts shown in Fig.~\ref{fig:fs}c with the corresponding theoretical prediction in Figs.~\ref{fig:fs}e--h and Fig.~\ref{fig:fs}i, respectively. The cuts at the center of the Brillouin zone using two perpendicular linear polarizations of the incident light clearly show that the large circular features at the $\Gamma$ point result from the intersection of holelike electronic dispersions with the Fermi level (indicated with white arrows in Figs.~\ref{fig:fs}e,f). At the very center of the Brillouin zone we further detect two more holelike electronic bands (dashed lines in Fig.~\ref{fig:fs}e), split by about $20\ \textrm{meV}$ (see Supplementary Information). The band top of the upper of these bands is located in the immediate vicinity of the Fermi level and produces the aforementioned localized intensity spot. In agreement with the experimental observations, the theoretical band structure predicts only bands with holelike dispersion at the $\Gamma$ point, as shown in Fig.~\ref{fig:fs}i. However, the theory predicts only three bands in contrast to the four bands clearly visible in the experiment (our detailed analysis of the experimental data based on the characteristic properties of the predicted electronic band structure, presented in the Supplementary Information, shows that the existence of five hole bands at the $\Gamma$ point is consistent with the data), indicating that some of the bands must be unrelated to the intrinsic bulk electronic structure of this compound.

\begin{figure}[!t]
\includegraphics[width=\textwidth/2]{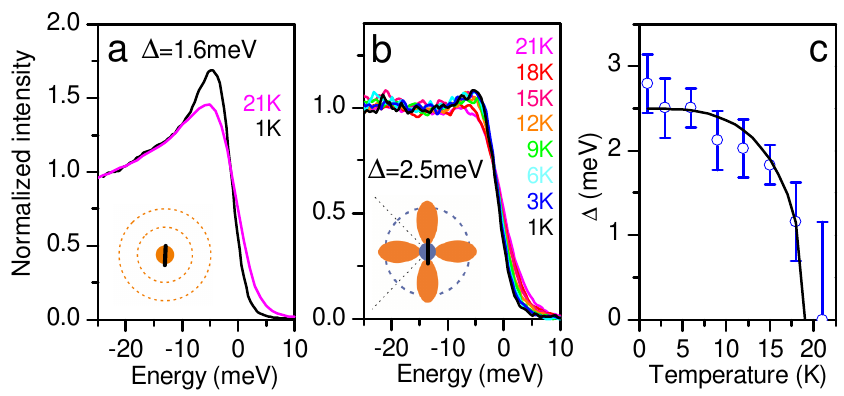}
\caption{\label{fig:sc}\textbf{Superconducting energy gap in the bulk low-energy electronic band structure of \sfcaopt.}~\textbf{a,b,} Energy-distribution curves in the {\it bulk} electronic structure at several temperatures in the superconducting (1K to 18K) and normal (21K) state integrated over a finite momentum range as shown by thick black lines in the respective insets. The energy of the incident radiation was tuned to $30\ \textrm{eV}$ and $35\ \textrm{eV}$ in \textbf{a} and \textbf{b}, respectively. The superconducting energy gaps extracted from a fit with the Dynes function (see discussion in the text) are indicated.~\textbf{c,} Temperature dependence of the superconducting energy gap in~\textbf{b} (empty circles) and a fit using the weak-coupling expression obtained in the Bardeen-Cooper-Schrieffer theory of superconductivity~\cite{Tinkham_superconductivity_1995_articlestyle}, with $\Delta(T=0)=2.5\ \textrm{meV}$ and $T_{\mathrm{c}}=19\ \textrm{K}$ (solid line).}
\end{figure}

The low-energy electronic structure in the corner of the Brillouin zone is much more complex. As can be seen in Fig.~\ref{fig:fs}h it features a small intensity spot at center of the propellerlike construct, which results from the bottom of a band with electronlike dispersion barely crossing the Fermi level from above (blue spots in Fig.~\ref{fig:fs}c and the dashed line in Fig.~\ref{fig:fs}h). Figure~\ref{fig:fs}h further shows another band with electronlike dispersion, whose bottom lies much deeper, about $135\ \textrm{meV}$ (see Supplementary Information), below the Fermi level and which produces the large circular intensity distribution in the corner of the Brillouin zone in Figs.~\ref{fig:fs}a--c. Figure~\ref{fig:fs}g, on the other hand, shows a very heavy band with holelike dispersion very close to the Fermi level along with additional coarsely perceived broad hole- and electronlike bands located at higher binding energies (dashed lines).

Quite remarkably, our {\it ab initio} calculations reveal the presence of all features observed in the cuts in Figs.~\ref{fig:fs}g,h and indicated as dashed lines, albeit {\it at a very different energy} below the Fermi level, as can be seen in Fig.~\ref{fig:fs}i. In order to understand what Fermi-surface topology this complex band structure generates, we plot constant-energy contours at three different energies in Figs.~\ref{fig:fs}j--l, as indicated in Fig.~\ref{fig:fs}i. As one can see, the typical nearly circular electronlike sheets in the corner of the Brillouin zone at the Fermi level predicted by the {\it ab initio} calculations (Fig.~\ref{fig:fs}d) evolve into mutually perpendicular ellipsoids of the same electronic character hybridized and split by the spin-orbital interaction in Fig.~\ref{fig:fs}j, then into two perpendicular linear intensity distributions in Fig.~\ref{fig:fs}k, and finally into four petallike structures of {\it holelike} character in Fig.~\ref{fig:fs}l. Importantly, this evolution involves a change of topology (transition from the ellipsoids formed by bands with electronlike dispersion into the petallike structures formed by those with holelike dispersion) over a very narrow energy range on the order of $100\ \textrm{meV}$ and thus may lead to a large variation in material properties in compounds with seemingly identical electronic structure.

From this comparison one can draw an important conclusion. It can be clearly seen that the deep electronlike band in Fig.~\ref{fig:fs}g does not generate a circular intensity distribution, as one could na\"ively expect from the $\Gamma$--$M$ cut in Fig.~\ref{fig:fs}i, but rather hybridizes with the heavy holelike band seen in Fig.~\ref{fig:fs}g to produce the petals of the propellerlike construct. This implies that the latter and the large circular feature produced by the deep electron band, indicated with white arrows in Fig.~\ref{fig:fs}h, are mutually incompatible and cannot coexist in the same band structure. Therefore, one of them must be extrinsic. This, together with the excess of bands with holelike dispersion at the $\Gamma$ point, implies the existence of two electronic structures contributing to the photoemission signal: one from the bulk of the material, the other one from its polar surface. 

In order to separate the contribution of the surface from the inherent bulk electronic structure of \sfcaopt, we have carried out a careful investigation of the surface of this material intentionally aged by repeated temperature cycling in a vacuum of $\sim10^{-9}\ \textrm{mbar}$. These results clearly indicate that the outer band with holelike dispersion at the $\Gamma$ point, generating the largest circular feature shown as an orange dashed circle in Fig.~\ref{fig:fs}c, is significantly suppressed in the aged material (see Supplementary Information), consistent with previous reports on related compounds~\cite{Shen_LaFePO_ElStr_2008,HaiYun3761,PhysRevLett.101.147003,Liu2009491,Lu2009452,PhysRevLett.105.027001,PhysRevB.82.075135,PhysRevB.82.104519,Yang2011460,PhysRevB.84.014504}. The extrinsic character of the second outermost hole band at the $\Gamma$ point follows from its incompatibility with the characteristic properties of the low-energy electronic band structure predicted by {\it ab initio} calculations (see Supplementary Information). The occurrence of such large circular features in the electronic structure of the 1111-type compounds is further in line with the predictions of {\it ab initio} slab calculations for an As-terminated surface~\cite{PhysRevB.81.155447}, which is expected to be significantly hole-doped. Thus the only bulk contribution to the photoemission signal at the center of the Brillouin zone comes from the inner band(s) in Figs.~\ref{fig:fs}e,f with holelike dispersion, which generates the localized intensity spot shown in orange in Fig.~\ref{fig:fs}c. 

Similarly, the deep electron band producing the large circular feature at the M point (blue dashed circle in Fig.~\ref{fig:fs}c) is completely absent in the aged material (see Supplementary Information) and must, therefore, be related to the surface states as well. The propellerlike structure, on the other hand, is unaffected by aging. This brings us to the central result of this study: the bulk electronic structure of the optimally doped \sfcaopt\ consists of several bands with holelike dispersion in the center of the Brillouin zone with the top of at least one of them in the immediate vicinity of the Fermi level as well as of a complex propellerlike construct in the corner of the Brillouin zone generated by a very shallow band with electronlike dispersion and a hybridized heavy holelike and a light (deeper) electronlike band. The inherent bulk photoemission intensity at the Fermi level can thus be entirely attributed to a set of band-edge singularities shown as blue and orange spots in Fig.~\ref{fig:fs}c. The corresponding bulk low-energy electronic structure is summarized schematically in Fig.~\ref{fig:bulksummary} (blue solid lines) along with the contribution to the photoemission signal from the polar surface of the sample (grey dashed lines).

\begin{figure}[!b]
\includegraphics[width=\textwidth/2]{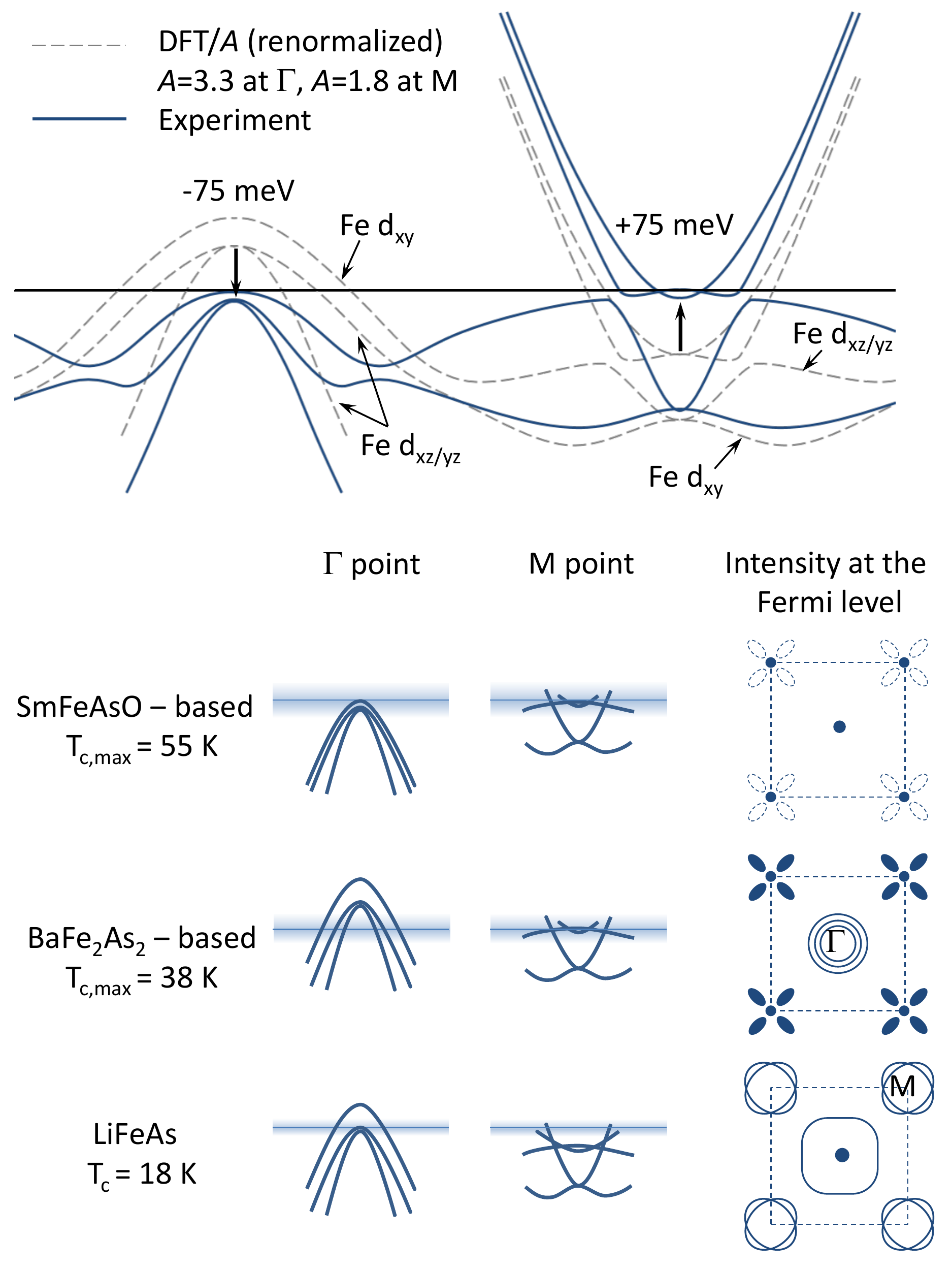}
\caption{\label{fig:elstrschematic}\textbf{Singularities in the electronic structure of various iron-based superconductors near the Fermi level.}~(top panel) Band-width renormalization by a factor of $\approx3.3$ at the $
\Gamma$ point and $\approx1.8$ at the $M$ point as well as band warping of order $75~\textrm{meV}$ (after renormalization) are required in order to reconcile the low-energy electronic band structure of 1111-type compounds obtained using {\it ab initio} calculations with the experimental observations on \sfcaopt\ reported in this work. These two interaction-induced effects lead to the formation of a singular Fermi surface in the majority of iron-based superconductors, as illustrated in the bottom panels. The number and intensity of these singularities appear to correlate positively with the magnitude of the maximum superconducting transition temperature found in these materials. The shaded areas in the bottom panels indicate schematically the respective energy windows around the Fermi level involved in the formation of the superconducting condensate. Similar singularities can be found in the Fermi-surface topology of the iron-chalcogenide ($A_{1-x}\textrm{Fe}_{2-y}\textrm{Se}_2$, $A$=K,~Rb,~Cs; FeSe; $\textrm{FeSe}_{1-x}\textrm{Te}_x$) superconductors~\cite{PhysRevB.88.134501} and monolayer FeSe on $\textrm{SrTiO}_3$ (Ref.~\onlinecite{2013NatMa12634T}).}
\end{figure}

Having established the bulk electronic structure of the optimally doped \sfcaopt\, we would like to note that superconductivity will develop on a very singular electronic landscape. Therefore, it is important to determine the momentum dependence of the superconducting energy gap throughout the Brillouin zone. To this end, we have carried out detailed high-resolution measurements of the temperature dependence of the photoemission intensity in the vicinity of both the $\Gamma$ and M points of the Brillouin zone. The results of these measurements are presented in Figs.~\ref{fig:sc}a,b in the form of energy-distribution curves (EDC) at two characteristic temperatures: in the normal state at $21\ \textrm{K}$ (magenta lines) and the superconducting state at $1\ \textrm{K}$ (black lines). In Fig.~\ref{fig:sc}b several intermediate temperatures are shown as well. The EDCs have been integrated over a finite momentum range (indicated schematically with black lines in the insets of the respective panels) to facilitate the extraction of the superconducting energy gap from experimental data with finite energy and momentum resolution, as described in Ref.~\onlinecite{PhysRevB.79.054517}. Even without modeling, these raw integrated data reveal the presence of a superconducting energy gap manifested in the shift of the leading-edge of the EDC curves at the Fermi wave vector $k_{\mathrm{F}}$. It is known, however, that the leading-edge shift deviates from the value of the superconducting energy gap due to finite experimental resolution~\cite{PhysRevB.79.054517}. Therefore, the value of the superconducting energy gap has been extracted by fitting the momentum-integrated EDCs in Fig.~\ref{fig:sc} with the Dynes function multiplied by the Fermi function and convolved with the response function:

\begin{equation}
\textrm{IEDC}(\omega)=\left[f(\omega,T)\left|Re\frac{\omega+i\Sigma^{\prime\prime}}{E}\right|\right]\otimes R_\omega(\delta E),
\end{equation}
where $E=\sqrt{(\omega+i\Sigma^{\prime\prime})^2-\Delta_{\mathbf{k}}^2}$, $\omega$ is the binding energy with reversed sign, $T$ is the temperature, $\Sigma^{\prime\prime}$ is the imaginary part of the self-energy, $\Delta_{\mathbf{k}}^2$ is the momentum-dependent superconducting energy gap, and $\delta E$ is the experimental resolution (see Ref.~\onlinecite{PhysRevB.79.054517}). The magnitude of the superconducting energy gap on the holelike band at the $\Gamma$ point has been found be $1.6\ \textrm{meV}$, while the center of the propellerlike structure exhibits a gap of $2.5\ \textrm{meV}$ with a clear BCS-type temperature dependence, as shown in Fig.~\ref{fig:sc}c. The corresponding gap ratios $2\Delta/(k_{\mathrm{B}}T_{\mathrm{c}})$ give $2$ and $3.16$, respectively, and are in a good agreement with the general trend found in the entire family of iron-based superconductors~\cite{PhysRevB.83.214520}. Thus the highly singular electronic band structure both at the $\Gamma$ and $M$ point of the Brillouin zone displays a clear superconducting energy gap. Finally, we have also detected the existence of a superconducting energy gap on the surface-related electronic bands (see Supplementary Information), which could be explained by the proximity effect of the superconducting bulk on the possibly non-superconducting surface.

The importance of the singular Fermi-surface topology observed in this work for superconductivity necessitates good microscopic understanding of its formation. Detailed comparison of the low-energy electronic structure of \sfcaopt\ in Fig.~\ref{fig:fs} with the predictions of the density-functional calculations reveals several important interaction-induced modifications, summarized in the top panel of Fig.~\ref{fig:elstrschematic}. First of all, the overall shape of the experimentally obtained band dispersions agrees very well with theoretical results if the latter are renormalized by a factor of $\approx3.3$ at the $\Gamma$ point and $\approx1.8$ at the $M$ point, comparable to the previous experimental observations on related materials~\cite{Shen_LaFePO_ElStr_2008,PhysRevB.84.014504,0034-4885-74-12-124512}. It has been suggested to originate in pronounced Hund's-coupling electron correlations observed in iron-based superconductors with a large number of experimental probes~\cite{YinHauleKotliar_HundsCouplingAndRenorm_2011,0953-8984-26-25-253203}. These correlations, in turn, can be expected and have indeed been found to be maximal near the perfect tetrahedral coordination of Fe and As ions and half filling of the Fe-d band~\cite{NakajimaUchidaOpticsBFA2014}. The overall band-structure renormalization, however, is insufficient to reconcile theory with experiment, at least in the case of \sfcaopt\ and several other iron-based compounds~\cite{PhysRevLett.105.067002,PhysRevLett.109.177001,PhysRevB.89.064514,PhysRevLett.110.167002}. In \sfcaopt\, the dispersion of the electronic states in the vicinity of the Fermi level shows an unusual band warping, which shifts the electronic structure in an orbital-dependent manner at the $\Gamma$ and M point of the Brillouin zone with respect to each other by about $150\ \textrm{meV}$ in the renormalized band structure, as indicated schematically in the top panel of Fig.~\ref{fig:elstrschematic}, and thereby pulls an unprecedented number of band-edge singularities to the immediate vicinity of the Fermi level. While the reliable identification of the microscopic mechanism behind such a band warping certainly requires further experimental efforts, it has been suggested that, in the presence of large particle-hole asymmetry generally found in the iron-based superconductors, a realistic interband interaction would lead to a relative shift of the coupled bands consistent with the experimentally observed values~\cite{Cappelluti2010S508}.

It is well-known that the superconducting properties of a material are defined by electronic states not only at the Fermi level but within an energy window of several superconducting energy gaps around it~\cite{RevModPhys.62.1027}. It is, therefore, clear that the concentration of a large density of states in the immediate vicinity of the Fermi level reported here must bear significantly upon the formation of the superconducting condensate. Our observation of the largest superconducting energy gap on the highly singular propellerlike construct in the corner of the Brillouin zone of \sfcaopt\ provides evidence for this statement. It has also been shown that very subtle changes in the Fermi surface of $\textrm{(Ba,K)}\textrm{Fe}_2\textrm{As}_2$ (disappearance of the center of the propellerlike structure at the M point of the Brillouin zone when going from $x=0.4$ to $x=0.6$ in doping) lead to a complete suppression of the superconducting gap on the outer sheet of the Fermi surface centered at the $\Gamma$ point~\cite{PhysRevB.86.165117}. Singularities in the bulk low-energy electronic structure near the Fermi level similar to those we observed in \sfcaopt\ in this work, but not all at once, have been found in other iron-based superconductors~\cite{PhysRevB.79.054517,PhysRevLett.105.067002,KanigelFeTeSe_vanHove_2012,PhysRevB.88.134501,PhysRevB.88.140505}. Quite importantly, these band-edge singularities are surprisingly robust with respect to doping, contrary to what one might expect based on a simple rigid-band--shift model of doping, as is evidenced by the observation of the propellerlike construct in $\textrm{(Ba,K)}\textrm{Fe}_2\textrm{As}_2$ at all doping levels from optimal doping to the extremely overdoped case of $\textrm{KFe}_2\textrm{As}_2$~\cite{Borisenko_BKFA_FS_NormState2009,PhysRevB.79.054517,PhysRevLett.103.047002}, despite very large additional charge donated by the dopants.

In this regard it is interesting to correlate the occurrence of singularities in the low-energy electronic structure of various iron-based superconductors with their respective superconducting transition temperatures near optimal doping. Such a comparison reveals a correlation between the number and intensity of band-structure singularities and the value of maximum $T_{\mathrm{c}}$ attainable in compounds of that particular class, demonstrated schematically in the bottom panel of Fig.~\ref{fig:elstrschematic} (see Refs.~\onlinecite{PhysRevB.79.054517,PhysRevLett.105.067002,KanigelFeTeSe_vanHove_2012,PhysRevB.88.134501,PhysRevB.88.140505}). A significant enhancement of the superconducting transition temperature in the vicinity of a band-structure singularity in iron-based and other layered superconductors has been suggested to result from the formation of Feshbach resonances~\cite{0953-2048-22-1-014004}. A similar effect in another strongly two-dimensional system, graphene with a superimposed electrical unidirectional superlattice potential, has recently been identified theoretically in the framework of the Bardeen-Cooper-Schrieffer theory of superconductivity~\cite{PhysRevB.89.195435}. Our observations clearly demonstrate that the singular experimental electronic band structure of optimally doped \sfcaopt\ (and many other iron-based superconductors) is strongly but comprehensibly modified from its theoretically predicted counterpart by sizable electronic interactions and urge that realistic band structure be adopted in favor of simple theoretical models used so far. They also suggest an interesting possible route towards the design of novel high-temperature superconductors by enhancing electronic interactions in a singular band structure.

\footnotesize
\mysection{Methods}
\par Angle-resolved photoemission measurements were performed using synchrotron radiation (``$1^3$--ARPES'' end-station at BESSY) within the range of photon
energies $20$-–$90\ \textrm{eV}$ and various polarizations on cleaved surfaces of high quality single crystals. The overall energy and angular resolution were $\sim 5\ \textrm{meV}$ ($8\ \textrm{meV}$) and $0.3^\circ$, respectively, for the low temperature measurements (FS mapping). The FS maps represent the momentum distribution of the intensity integrated within a $10\ \textrm{meV}$ window at the Fermi level. To equalize the intensity over different Brillouin zones the maps were normalised at each $\mathbf{k}$ point to the total recorded EDC intensity.
\par High-quality single crystals of superconducting $\textrm{SmFe}_{0.92}\textrm{Co}_{0.08}\textrm{AsO}$ with masses of a few micrograms were synthesized by the high-pressure high-temperature cubic anvil technique and were characterized by x-ray diffraction, transport and magnetization measurements~\cite{PhysRevB.86.214509}. The latter revealed a superconducting transition temperature of about $16\ \textrm{K}$. The width of the superconducting transition was found to be less than $1\ \textrm{K}$, indicating very high homogeneity of the investigated samples.
\par Band structure calculations were performed for the experimental crystal structure of SmFe$_{0.92}$Co$_{0.08}$AsO~\cite{PhysRevB.86.214509} using the PY~LMTO computer code \cite{Yaresko_LDA_2004_articlestyle}. Calculations based on the local density approximation (LDA) tend to put partially filled 4$f$ states of lanthanides at the Fermi level. In order to avoid this in the present calculations a Sm ion was replaced by La. We have verified that the band dispersions in the vicinity of the Fermi level calculated for LaFeAsO are very close to dispersions obtained for SmFeAsO with the same atomic positions provided that Sm~$4f^{5}$ electrons treated as localized quasi-core states and their exchange splitting is neglected. The effect of Co doping was simulated by using the virtual crystal approximation.


\mysection{Acknowledgements}
\noindent This project was supported by the German Science Foundation under Grants No. BO 1912/2-2 within SPP 1458. Work at ETH was partially supported by the SNSF and the NCCR program MaNEP. A.C. acknowledges financial support by the
Alexander von Humboldt foundation. We would like to thank D. V. Evtushinsky for useful discussions regarding data analysis. 
\mysection{Author contributions}
\noindent S.T., V.B.Z., and S.V.B. carried out the experiments. A.C., S.T., and S.V.B. analyzed the data. A.C. and  S.V.B. wrote the manuscript. N.D.Z. carried out the sample growth and characterization. A.N.D. carried out {\it ab initio} calculations. S.V.B., B.Bg., and B.B. supervised the project. All authors discussed the results and reviewed the manuscript.
\mysection{Additional information}
\noindent {\bf Supplementary Information} accompanies this paper on www.nature.com/scientificreports.\\
\noindent {\bf Competing financial interests:} The authors declare no competing financial interests.\\
\noindent Correspondence should be addressed to A.C. (acharnukha@ucsd.edu) and S.V.B. (s.borysenko@ifw-dresden.de) Requests for materials should be addressed to N.D.Z. (zhigadlo@phys.ethz.ch) and B.Bg. (batlogg@solid.phys.ethz.ch).
\eject\newpage

\setcounter{figure}{0}
\renewcommand{\bibnumfmt}[1]{#1.}
\renewcommand{\bibnumfmt}[1]{S#1.}
\renewcommand\thefigure{S\arabic{figure}}
\renewcommand{\cite}[1]{[S\citenum{#1}]}

\thispagestyle{plainfancy}

\fontfamily{helvet}\fontseries{bf}\selectfont
\mathversion{bold}
\begin{widetext}
\begin{figure}
\vskip0pt\noindent\hskip-0pt
\hrule width\headwidth height\headrulewidth \vskip-\headrulewidth
\hbox{}\vspace{4pt}
\hbox{}\noindent\vskip10pt
\begin{center}
\huge\sffamily\textbf{Supplementary information}
\end{center}
\end{figure}
\end{widetext}
\normalfont\normalsize
\section{Low-energy electronic structure of $\textbf{SmFe}_{\mathbf{0.92}}\textbf{Co}_{\mathbf{0.08}}\textbf{AsO}$ near the $\Gamma$ point}
\par
\begin{figure}[h]
\includegraphics[width=\textwidth]{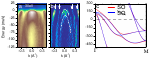}
\caption{\label{fig:elstrgamma}\textbf{Band splitting in the low-energy electronic structure of \sfcaopt\ near the $\Gamma$ point}~\textbf{a} Momentum-energy cut of the photoemission signal in the $\Gamma$--M direction of the Brillouin zone obtained using an incident photon energy of $30\ \textrm{eV}$.~\textbf{b} Second derivative of the intensity distribution shown in panel~\textbf{a} taken along the momentum axis. White arrows indicate bands clearly crossing the Fermi level, while the white dashed lines indicate those terminating in its immediate vicinity.~\textbf{c} Theoretically predicted low-energy electronic structure along the $\Gamma$--M direction in the Brillouin zone, with (red lines) and without (blue lines) the spin-orbit coupling taken into account.}
\end{figure}

 In order to confirm the existence of {\it two} electronic bands terminating in the immediate vicinity of the $\Gamma$ point but split by an amount of approximately $20\ \textrm{meV}$, as stated in the main text, we plot in Fig.~\ref{fig:elstrgamma} the distribution of the photoemission intensity in an energy-momentum cut along the $\Gamma$--M direction of the Brillouin zone as well as the second derivative of this signal taken along the momentum axis. Already the raw data in Fig.~\ref{fig:elstrgamma}a clearly indicate the presence of four bands with holelike electronic dispersion around the $\Gamma$, two of which cross the Fermi level (white arrows in Fig.~\ref{fig:elstrgamma}b), while the other two terminate in its immediate vicinity (white dashed lines in Fig.~\ref{fig:elstrgamma}b). The splitting between the latter is $\approx22\ \textrm{meV}$. It is much smaller than its theoretically predicted counterpart (about $80~\textrm{meV}$, see Fig.1i of the main text and Fig.~\ref{fig:elstrgamma}c). However, if one assumes that this splitting is renormalized along with the underlying band structure then its renormalization ($80/22=3.6$) is fully consistent with the renormalization of the hole bands at the G point (a factor of about $3.3$, see main text and Fig.3 therein). Figure~\ref{fig:elstrgamma}c further clearly shows that the lower two hole bands, degenerate by symmetry in the absence of the spin-orbit coupling, become split when the spin-orbit coupling is taken into account. However, this splitting is significantly smaller than that between the upper two bands and could not be resolved in our measurements.
\section{Effect of surface aging on the low-energy electronic structure of $\textbf{SmFe}_{\mathbf{0.92}}\textbf{Co}_{\mathbf{0.08}}\textbf{AsO}$}
\par
\begin{figure}[h]
\includegraphics[width=\textwidth]{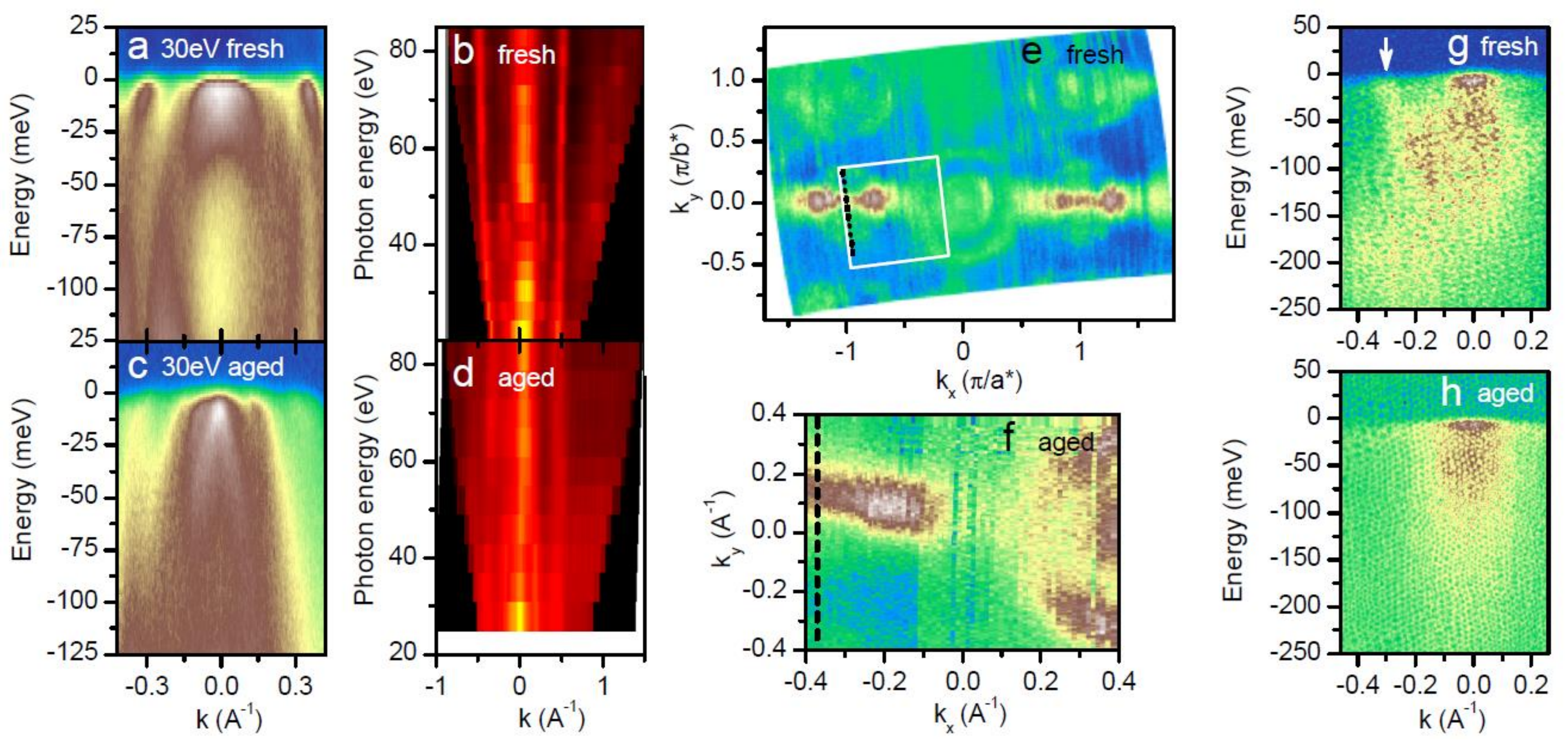}
\caption{\label{fig:aging}\textbf{Effect of surface aging on the low-energy electronic structure of \sfcaopt}~\textbf{a,c,} Band dispersions in the vicinity of the $\Gamma$ point obtained using incident radiation polarized within the plane of incidence (horizontal polarization) before~(\textbf{a}) and after~(\textbf{c}) aging, respectively.~\textbf{b,d} Dependence of the photoemission intensity in the vicinity of the Fermi level along the $\Gamma$-M high-symmetry direction on the energy of the incident radiation, revealing the $k_{\mathrm{z}}$ dispersion of the holelike bands near the $\Gamma$ point, before and after aging, respectively.~\textbf{e,} Constant-energy map obtained on the fresh sample, from Fig.1b of the main text.~\textbf{f,} The same for the aged sample in a narrow momentum region, indicated with a white rectangle in panel~\textbf{e}. This constant-energy map was obtained by integrating the photoemission intensity in a small energy window of $10~\textrm{meV}$ around the Fermi level recorded at $T=1~\textrm{K}$ using photons with an excitation energy of 35 eV linearly polarized perpendicular to the plane of incidence.~\textbf{g,h} Energy-momentum cuts across the propellerlike structure along the black dashed lines in panels~\textbf{e,f} before and after aging, respectively. The white arrow in panel~\textbf{g} indicates the deep electron band giving rise to the large circular feature at the M point in panel~\textbf{e}.}
\end{figure}

 To confirm the assignment of the band dispersions producing large circular intensity distributions near the $\Gamma$ and M point in the photoemission maps of Fig.1a,b to the polar surface, we have subjected the sample surface of \sfcaopt\ to intentional aging via temperature cycling between $300$ and $1\ \textrm{K}$. The results of this experiment are summarized in Fig.~\ref{fig:aging}. Panels~\ref{fig:aging}a,c demonstrate the effect of aging on the low-energy electronic structure in the vicinity of the $\Gamma$ point. While the freshly cleaved sample (Fig.~\ref{fig:aging}a) clearly shows three holelike bands of comparable intensity crossing or approaching the Fermi level, in the aged material (Fig.~\ref{fig:aging}c) the large outer dispersion is suppressed, strongly supporting the assignment of this band to the surface-related electronic structure in the main text. The same conclusion can be drawn from the comparison of the photon-energy--dependence of the photoemission intensity near the Fermi level before and after aging shown in Figs.~\ref{fig:aging}b,~d, respectively. The two outer features are very two-dimensional (have negligible $k_{\mathrm{z}}$ dispersion) and the outermost one is strongly suppressed in the aged sample, with the central bulk-related feature virtually unchanged. While the suppression of the intensity of the second outermost band is quite small, our analysis of the characteristic properties of the electronic band structure of \sfcaopt\ clearly indicates that this band is incompatible with the bulk and must be extrinsic as well.
Surface aging has a strong effect on the low energy electronic structure at the M point of the Brillouin zone as well. While the photoemission intensity map in the fresh sample in Fig.~\ref{fig:aging}e clearly shows the presence of a large circular feature at the M point, this latter is entirely suppressed in the aged sample, as shown in Fig.~\ref{fig:aging}f. This effect of aging can be seen even more clearly in the energy-momentum cuts across the propellerlike structure shown in Figs.~\ref{fig:aging}g,~h (taken along the dashed lines in Figs.~\ref{fig:aging}e,~f, respectively): the deep electron band giving rise to the large circular feature at the M point (indicated with the white arrow in Fig.~\ref{fig:aging}g) vanishes entirely in the aged sample, whereas the bands contributing to the propellerlike structure remain unaffected. It must, therefore, be concluded that the propellerlike construct is a part of the bulk electronic structure, while the deep electron band producing the large circular feature in Fig.~\ref{fig:aging}e must have a surface-related character.
\section{Analysis of the surface-related electron band at the M point}
\begin{figure}[h]
\includegraphics[width=\textwidth/2]{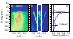}
\caption{\label{fig:surfelpocket}\textbf{Low-energy electronic structure of \sfcaopt\ at the M point.}~\textbf{a,} Energy-momentum cut at the M point perpendicular to the $\Gamma$--M direction obtained using photons with an excitation energy of $80~\textrm{eV}$ linearly polarized perpendicular to the plane of incidence (vertical polarization).~\textbf{b,} Second derivative of the data in~\textbf{a} with respect to momentum.~\textbf{c,} Energy-distribution curve averaged in the momentum window shown as a white rectangle in panel~\textbf{b}. Horizontal dashed lines indicate the location of the Fermi energy and the bottom of the deep electron band evident in panels ~\textbf{a,b}.}
\end{figure}

We now turn to the detailed analysis of the deep surface-related electron band producing the large circular intensity distribution at the M point in Fig.~\ref{fig:aging}g. Figure~\ref{fig:surfelpocket}a shows an energy-momentum cut along the left edge of the white rectangle in Fig.~\ref{fig:aging}g. A very shallow and a much deeper electron band producing the localized intensity spot and a large circular feature near the Fermi level, respectively, are clearly visible. In order to determine the location of the bottom of the deep surface-related electron band, we have taken the second derivative of the experimental data in Fig.~\ref{fig:surfelpocket}a with respect to momentum, shown in Fig.~\ref{fig:surfelpocket}b. The analysis of the energy-distribution curve averaged within a finite momentum window (indicated by a white rectangle in Fig.~\ref{fig:surfelpocket}b) in Fig.~\ref{fig:surfelpocket}c indicates that the band bottom is located near $135~\textrm{meV}$ below the Fermi level.
\section{Separation of the surface-related bands from the bulk based on the characteristic properties of the electronic structure}
In this section we would like to utilize the characteristic properties of the electronic band structure, well-established from {\it ab initio} density-functional calculations on iron pnictides in general and the 1111-type materials in particular, to prove the surface-related character of the second outermost band at the $\Gamma$ point and shed further light onto the low energy bulk electronic structure in the same region of the Brillouin zone.
\begin{figure}[h]
\includegraphics[width=\textwidth]{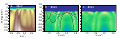}
\caption{\label{fig:connection}\textbf{Continuity of the low-energy electronic structure in \sfcaopt.}~\textbf{a,} Energy-momentum cut from Fig.~\ref{fig:aging}b in a larger energy and momentum window.~\textbf{b,} Energy-momentum cut along the $\Gamma$--M direction from the data in Fig.1b of the main text. Black dashed lines represent schematically the bulk electronic band structure expected from the {\it ab initio} calculation shown in Fig.1i of the main text. White dashed lines indicate the surface-related bands.~\textbf{c,} Same as panel~\textbf{b} but without dashed lines to demonstrate the features in the raw data.}
\end{figure}
First of all, based on the electronic structure shown in Fig.1i of the main text, we would like to point out that irrespective of the assignment of various bands observed in the experiment to the surface or bulk, several conditions must be met: a) the heavy hole band at the M point must connect to one of the three bands at the $\Gamma$ point closest to the Fermi level; b) the bottom of the deep electron pocket at the M point must connect to another one of the three bands at the $\Gamma$ point closest to the Fermi level; c) the third band of the three closest to the Fermi level does not connect to the propellerlike construct at the M point and disperses towards higher binding energies in the limited energy window shown in Fig.1i of the main text; d) the deeper two of the three aforementioned bands at the $\Gamma$ point are quasi-degenerate at the center of the Brillouin zone, the other one is split off them by spin-orbit coupling.

Based on these observations we now show that neither of the two outermost bands at the $\Gamma$ point can belong to the same band structure as the inner hole bands terminating in the immediate vicinity of the Fermi level and the propellerlike construct at the M point. Figure~\ref{fig:connection}a shows that the two outer bands at the $\Gamma$ point disperse quasi-quadratically towards high binding energies without signs of rounding off towards the edges of the Brillouin zone, whereas the heavy hole band at the M point (visible in the top left corner of Fig.~\ref{fig:connection}a and, more clearly, in Figs.~\ref{fig:connection}b,c) connects to one of the bands at $\Gamma$ within $50~\textrm{meV}$ below the Fermi level. Given that the propellerlike structure is not affected by aging (as opposed to the deep electron band giving rise to the large circular feature at the M point in Fig.~\ref{fig:aging}g, which is clearly suppressed by aging) one must conclude that all the bands producing large circular features in Fig.1a,b of the main text are surface-related.
The only remaining issue is the lack of the third hole band in the low energy electronic structure at the $\Gamma$ point, expected from the {\it ab initio} calculations (see Fig.1i of the main text). Indeed, up to now we have only shown the existence of two hole bands terminating in the immediate vicinity of the Fermi level (see Fig.1e and Fig.~\ref{fig:elstrgamma}), with a splitting of about $25~\textrm{meV}$. Quite interestingly, the same analysis as above allows one to maintain that our data are, in fact, consistent with the presence of all three hole bands at the $\Gamma$ point, shown in Fig.1i of the main text. This can be seen as follows: the top bulk hole band at the $\Gamma$ point must connect to the heavy hole band at the M point. The splitting between the other two hole bands at the $\Gamma$ point due to the spin-orbit coupling is expected to be quite small (see Fig.~\ref{fig:elstrgamma}c) and could not be resolved in our measurements. However, one of them must connect to the bottom of the propellerlike structure at M within about $150$--$175~\textrm{meV}$ below the Fermi level and thus must round off and cannot continue dispersing to higher binding energies. The only of the considered here three hole bands at the $\Gamma$ point that can is the innermost hole band, as can be seen in Fig.1i of the main text. Figures~\ref{fig:connection}b,c clearly show the presence of a hole band dispersing quasi-quadratically down to at least $300~\textrm{meV}$. One must, therefore, conclude that all three hole bands predicted by ab inito calculations contribute to the photoemission intensity near the $\Gamma$ point.
\section{Superconducting energy gap on the surface-related photoemission features near the $\Gamma$ point}
\begin{figure}[h]
\includegraphics[width=\textwidth/2]{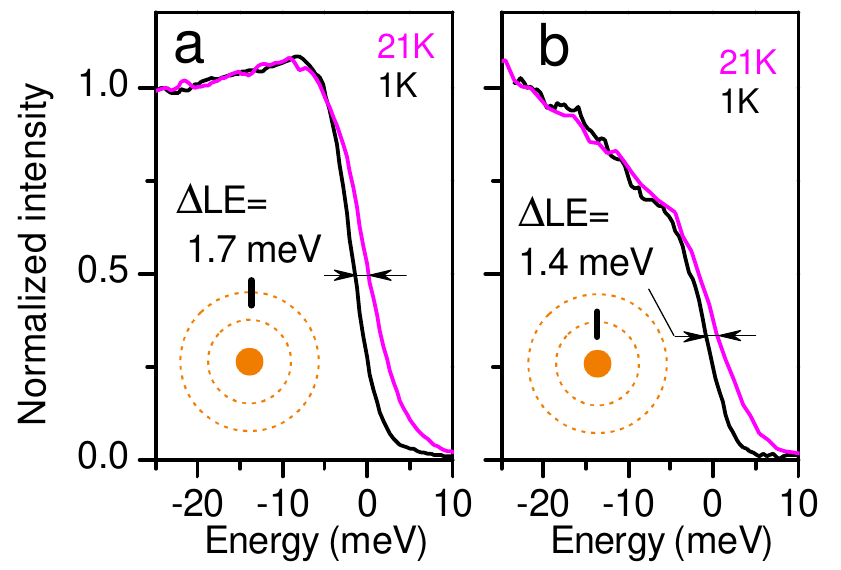}
\caption{\label{fig:scsuppl}\textbf{Superconducting energy gap on the surface-related bands of \sfcaopt.}~\textbf{a,b,} Energy-distribution curves in the surface-related electronic structure in the superconducting ($1\ \textrm{K}$) and normal ($21\ \textrm{K}$) state obtained at an energy of the incident radiation of $30\ \textrm{eV}$ and integrated over a finite momentum range as shown by thick black lines in the respective insets. The shift of the leading edge indicates the presence of a superconducting energy gap at low temperatures.}
\end{figure}

In order to investigate the effect of superconductivity on the surface-related features (the two outer band dispersions) near the $\Gamma$ point, we have studied the temperature dependence of these features deep in the superconducting and normal state at an energy of the incident radiation of $30\ \textrm{eV}$. The results of these measurements are presented in Fig.~\ref{fig:scsuppl} in the form of energy-distribution curves (EDC) at two characteristic temperatures: in the normal state at $21\ \textrm{K}$ (magenta lines) and the superconducting state at $1\ \textrm{K}$ (black lines). The EDCs have been integrated over a finite momentum range (indicated schematically with black lines in the insets of the respective panels). Both EDCs show a very clear shift of the leading edge on the order of $1.5\ \textrm{meV}$ indicating the presence of a superconducting energy gap in the corresponding electronic dispersions. Since the surface-related bands shown in Fig.~\ref{fig:scsuppl} do not exhibit any coherence peak in the superconducting state, the magnitude of the superconducting energy gap cannot be extracted reliably through fitting. Therefore, only the leading-edge shift is indicated in the panels as the lower bound of the superconducting energy gap on these bands.
\end{document}